\documentclass[twocolumn,showpacs,amsmath,amssymb]{revtex4}
\usepackage{graphicx}

\begin{document}

\title{Exact Mapping from Singular Value Spectrum of Fractal Images to Entanglement Spectrum of One-Dimensional Quantum Systems}

\author{Ching Hua Lee${}^{a}$}
\author{Yuki Yamada${}^{b}$}
\author{Tatsuya Kumamoto${}^{b}$}
\author{Hiroaki Matsueda${}^{b}$}
\affiliation{
${}^{a}$Department of Physics, Stanford University, CA 94305, USA \\
${}^{b}$Sendai National College of Technology, Sendai 989-3128, Japan
}

\date{\today}
\begin{abstract}
We examine the snapshot entropy of general fractal images defined by their singular values. Remarkably, the singular values for a large class of fractals are in exact correspondence with the entanglement spectrum of free fermions in one dimension. These fermions allow for a holographic interpretation of the logarithmic scaling of the snapshot entropy, which is in agreement with the Calabrese-Cardy formula. However, the coarse-grained entropy exhibits a linear scaling due to the degeneracy of the spectrum, in contrast with the logarithmic scaling behavior in one-dimensional quantum near-critical systems.
\end{abstract}
\pacs{05.10.Cc, 07.05.Pj, 11.25.Hf, 11.25.Tq, 89.70.Cf}
\maketitle

The study of quantum entanglement has continuously attracted enormous attention. The most important aspect concerns the scaling relation of the entanglement entropy in various quantum systems. Well-known scaling relations are the area law, the Calabrese-Cardy formula~\cite{Holzhey,Calabrese}, and the finite-entanglement scaling (we call this as finite-$\chi$ scaling for simplicity)~\cite{Tagliacozzo,Pollmann}. Since the latter two formulae enable us to estimate the central charge, the entropy is a poweful tool for the study of critical phenomena.

On the other hand, quantum entanglement is also deeply intertwined with the theme of holography. Quantum entanglement holds the key to surprising holographic correspondence between completely different physical systems. Two important manifestations are the anti-de Sitter space / conformal field theory correspondence in string theory and the multiscale entanglement renormalization ansatz in statistical physics. In order to compare between different systems, a crucial factor is the amount of information behind these systems, not their detailed physical properties. Thus, the entropy plays a central role in the comparison. Since any classical system does not have entanglement, we must reconsider the meaning of entanglement entropy in the classical side, if there exists possible classical representation of entanglement. In the conformal field theory language, the entanglement entropy is logarithm of two point correlation function of scaling operators, and this indicates the entropy contains the information from the physics at different length scales. Then, it should be possible to encode these degrees of freedom in an emergent space with an additional dimension. One of the solutions for the encoding is the so-called Ryu-Takayanagi formula~\cite{RT}. 

Furthermore, the Suzuki-Trotter decomposition is also a well-known quantum-classical correspondence. A typical example is transformation of the transverse-field Ising chain into the anisotropic two-dimensional (2D) classical spin model. One of the authors has found that the entropy of the spin snapshot in the classical system corresponds to the holographic entanglemenet entropy of the original quantum 1D system~\cite{Matsueda}. Remarkably, the snapshot entropy contains an equivalent amount of information as the Calabrese-Cardy and finite-$\chi$ scaling formulae combined. There, the singular value decomposition (SVD) of the snapshot data is a quite essential procedure for defining the snapshot entropy. A discretized holographic space emerges from this decomposition in the sense that the decomposed data have actually their own length scales. The holographic entropy scaling looks at such sequence of different length-scale information.

The purpose of this letter is to derive the exact mapping of the singular values of fractal images to the entanglement spectrum in 1D quantum systems. This is because the snapshot entropy, in spite of its potential applicability, is not still a well-defined quantity in some sense, and thus it's quite necessary to connect it to more physical systems. We find that the mapping produces free fermionic and more exotic particles, and then the discrete scaling symmetry is mapped onto the degeneracy of their entanglement spectra. In this paper, we focus on the mapping onto free fermions. We would like to also discuss about the finite-$\chi$ scaling. We will prove that the finite-$\chi$ scaling should be linear in the fractal cases in contrast to $\ln\chi$ in quantum near-critical 1D systems. Compared to the Calabrese-Cardy formula, the finite-$\chi$ scaling depends more delicately on the entanglement spectrum, and is thus a good measure to detect the difference between simple scale and full conformal symmetries.

We start from matrix data of a $L\times L$ image $M(x,y)$ in which each element takes an integer value ranging from $0$ to $1$. The value $0$ denotes a white pixel, and $1$ black. We apply SVD to $M(x,y)$ as
\begin{eqnarray}
M(x,y)&=&\sum_{l=1}^{L}M^{(l)}(x,y), \\
M^{(l)}(x,y)&=&U_{l}(x)\sqrt{\Lambda_{l}}V_{l}(y),
\end{eqnarray}
where $\Lambda_{l}$ denote the singular values and $U_{l}(x)$ and $V_{l}(y)$ are column unitary matrices. We normalize the singular values as $\lambda_{l}=\Lambda_{l}/\sum_{l}\Lambda_{l}$, and we arrange the order of $\lambda_{l}$ so that $\lambda_{1}\ge\lambda_{2}\ge\cdots$. Since all of the singular values are non-negative ones, this normalization leads to a probability distribution. We also define the coarse-grained snapshot and entropy with $\chi$ states kept as
\begin{eqnarray}
M_{\chi}&=&\sum_{l=1}^{\chi}M^{(l)}(x,y) , \\
S_{\chi}&=&-\sum_{l=1}^{\chi}\lambda_{l}\ln\lambda_{l}. \label{entropy}
\end{eqnarray}
Here we abbreviate the full entropy $S_{L}$ as $S$. The entropy measures the entanglement (take care about this terminology, this is not quantum entanglement) between the vertical and horizontal components $U_{l}(x)$ and $V_{l}(y)$. The singular values are the eigenvalues of the density matrices defined by
\begin{eqnarray}
\rho_{X}(x,x^{\prime})&=&\sum_{y}M(x,y)M(x^{\prime},y)=\sum_{l}U_{l}(x)\Lambda_{l}U_{l}(x^{\prime}), \nonumber \\
&& \\
\rho_{Y}(y,y^{\prime})&=&\sum_{x}M(x,y)M(x,y^{\prime})=\sum_{l}V_{l}(y)\Lambda_{l}V_{l}(y^{\prime}). \nonumber \\
\end{eqnarray}
Their eigenvalues are the same. This property is similar to quantum entanglement between two subsystems.

\begin{figure}[htbp]
\begin{center}
\includegraphics[width=7cm]{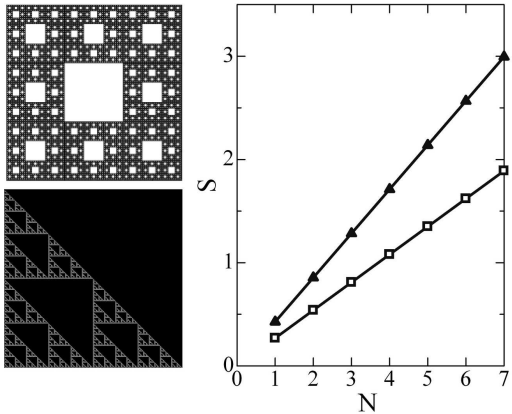}
\end{center}
\caption{(left upper panel) Sierpinski carpet, (left lower panel) Sierpinski triangle, (right panel) full snapshot entropy as a function of the fractal level $N$. The open squares (filled triangles) represent $S$ for the carpet (triangle).}
\label{fig01}
\end{figure}

As typical examples, we show a Sierpinski carpet and a Sierpinski triangle in Fig.~\ref{fig01}. In the carpet case, we first devide a whole area into nine blocks, and occupy the center block by white color. In the next step, the eight black blocks are respectively devided into nine blocks, and each center of them is occupied by white color. We repeat this process until the minimal length scale comes to our finite system size $L$. We call this fractal as white-centered one. The fractal dimension $D$ of this image is calculated as $D=\ln 8 / \ln 3=1.893$.

A method of general mapping is quite simple, but yet very powerful. Let us start with the $h\times h$ unit cell of a given fractal. In the Sierpinski carpet and triangle, the unit cells are respectively given by the following matrices
\begin{eqnarray}
H=\left(\begin{array}{ccc}1&1&1\\ 1&0&1\\ 1&1&1\end{array}\right)\; , \; H=\left(\begin{array}{cc}1&0\\0.9&1.5\end{array}\right).
\end{eqnarray}
The entries of the fractals can be continuously tuned to obtain the desired luminance, like in the case of the Sierpinski triangle shown. Now, we construct a $h^{N}\times h^{N}$ fractal matrix $M$ ($L=h^{N}$). This can be easily done by taking the tensor product of $N$ copies $M=H\otimes H\otimes\cdots\otimes H$. Usually if this factorization occurs in a quantum state, such a state is called pure state. On the other hand, due to its multiple scales of self-similarity, a fractal image cannot be decomposed into the direct product form of the vertical and horizontal components by SVD, and in that sense the fractal is entangled. It is thus interesting that the product form represents such entangled structure. If the eigenvalues of $H$ are $\gamma_{1},\gamma_{2},\cdots,\gamma_{h}$, the non-zero eigenvalues of $M$ are given by
\begin{eqnarray}
\Gamma_{\mbox{\boldmath $a$}}=\prod_{j=1}^{N}\gamma_{a_{j}},
\end{eqnarray}
where each $a_{j}$ takes values from $1$ to $r={\rm rank}H$ ($r\le h$). In the next paragraph, we show that the system is transformed into free fermions when $r=2$ as in the cases of Sierpinski carpet and triangle. This fact does not depend on $h$. The condition, $r=2$, is true for a large class of fractals. We normalize $\Gamma_{\mbox{\boldmath $a$}}$ as $\lambda_{\mbox{\boldmath $a$}}=\Gamma_{\mbox{\boldmath $a$}}^{2}/\sum_{\mbox{\boldmath $a$}}\Gamma_{\mbox{\boldmath $a$}}^{2}$, and we write $c_{k}=\gamma_{k}^{2}/\sum_{k=1}^{r}\gamma_{k}^{2}$. Then, the entropy is given by
\begin{eqnarray}
S=-\sum_{\mbox{\boldmath $a$}}\lambda_{\mbox{\boldmath $a$}}\ln\lambda_{\mbox{\boldmath $a$}}=-N\sum_{k=1}^{r}c_{k}\ln c_{k}\propto \ln L. \label{generalS}
\end{eqnarray}
This result guarantees the logarithmic entropy formula for any $r$ and $h$ values. Here, it should be noted that there are fractals that cannot be written in the tensor-product form, even though they still look self-similar. One explicit counterexample is the black-centered Sierpinski carpet. The unit cell $H^{\prime}$ is given by
\begin{eqnarray}
H^{\prime}=\left(\begin{array}{ccc}0&0&0\\ 0&1&0\\ 0&0&0\end{array}\right)=B-H, \;\; B=\left(\begin{array}{ccc}1&1&1\\ 1&1&1\\ 1&1&1\end{array}\right),
\end{eqnarray}
where $H$ is the unit cell of the white-centered carpet. The matrix $B$ disturbs the factorized form. In general, the entropy will not scale logarithmically when $B$ and $H$ do not commute, as we shall discuss at length in future works. Our entropy is asymmetric with respect to the exchange of white and black pixels.

Let us take a more precise look at the white-centered Sierpinski carpet. In this case, $H$ has two non-zero eigenvalues $\gamma_{\pm}=1\pm\sqrt{3}$ ($r=2$). The total number of the non-zero eigenvalues of $M$ is then $2^{N}$. The independent eigenvalues are given by
\begin{eqnarray}
\Gamma_{j}=\gamma_{+}^{j}\gamma_{-}^{N-j},
\end{eqnarray}
with $j$ running from $0$ to $N$, and the degeneracy is represented by the binomial coefficient $\alpha_{j}=N!/j!(N-j)!$ with $\sum_{j=0}^{N}\alpha_{j}=2^{N}$. The characteristic polynomial of $M$ can be represented by 
\begin{eqnarray}
p(z)=\left|zI-M\right|=z^{3^{N}-2^{N}}\prod_{j=0}^{N}\left(z-\Gamma_{j}\right)^{\alpha_{j}},
\end{eqnarray}
where $I$ is the unit matrix of $L\times L$. Then, the normalized singular values of $M$ are given by
\begin{eqnarray}
\bar{\lambda}_{j}=\frac{\Gamma_{j}^{2}}{\sum_{j=0}^{N}\alpha_{j}\Gamma_{j}^{2}}=\left(\frac{1}{2}+\frac{\sqrt{3}}{4}\right)^{j}\left(\frac{1}{2}-\frac{\sqrt{3}}{4}\right)^{N-j}. \label{ss}
\end{eqnarray}
where we have done re-labeling of the index so that $\bar{\lambda}_{N}=\lambda_{1}$, $\bar{\lambda}_{N-1}=\{\lambda_{2},...,\lambda_{N+1}\}$, etc.

On the other hand, let us consider $N$ sites of an infinite free-fermion chain. According to Refs.~\cite{Peschel,Vidal,Calabrese2}, the $2^{N}$ eigenvalues of the reduced density matrix are given by the products
\begin{eqnarray}
\lambda=\prod_{k\in A}\nu_{k}\prod_{k\in B}\left(1-\nu_{k}\right). \label{ff}
\end{eqnarray}
where we consider that $A$ is a subset of $\{1,2,...,N\}$ and $B$ is th rest of the subset. Here $\nu_{k}$ correspond to the eigenvalues of a $N\times N$ single-particle correlator matrix $C_{ij}={\rm Tr}(\rho c_{j}^{\dagger}c_{j})$, when the reduced density matrix is given by $\rho\propto\exp\left(-\sum_{1\le i,j\le l}h_{ij}c_{i}^{\dagger}c_{j}\right)$ with a $N\times N$ matrix $h=\ln[(I-C)C^{-1}]$. The inverse transformation actually gives $\nu_{k}=(e^{\epsilon_{k}}+1)^{-1}$ with the entanglement energy $\epsilon_{k}$.

Comparing Eq.~(\ref{ss}) with Eq.~(\ref{ff}), we clearly observe one-to-one correspondence between them
\begin{eqnarray}
\nu_{k}=\frac{1}{2}+\frac{\sqrt{3}}{4}.
\end{eqnarray}
It is straightforward to write down the entropy formula from Eq.~(\ref{generalS}) as
\begin{eqnarray}
S &=& -\sum_{k=0}^{N}\left\{\nu_{k}\ln\nu_{k}+(1-\nu_{k})\ln(1-\nu_{k})\right\} \nonumber \\
&=& 0.245775\ln L. \label{log}
\end{eqnarray}
Therefore, the snapshot entropy should obey this logarithmic scaling, and the scaling agrees with the Calabrese-Cardy formula
\begin{eqnarray}
S=\frac{c}{3}\ln L,
\end{eqnarray}
where $c$ is the central charge. In Fig.~\ref{fig01}, we present numerical data of the full snapshot entropy $S$ as a function of the fractal level $N$. For the Sierpinski carpet, we actually find that the entropy perfectly matches with Eq.~(\ref{log}) ($S=0.245775\ln L=0.270011N$).

\begin{figure}[htbp]
\begin{center}
\includegraphics[width=7.5cm]{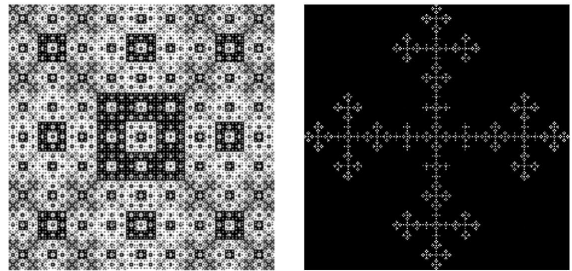}
\end{center}
\caption{Induced patterns: (left) $\delta=-4$, (right) $\delta=-\sqrt{3}$.}
\label{fig02}
\end{figure}

The exact result provides us with several important pieces of information. Firstly, the fractal level $N$ is the degeneracy of the electron's spectrum. In other words, the scaling symmetry is manifested in the degeneracy of the fermion system. Secondly, the quality of the fractal image is measured by how close the value $\gamma_{-}^{2}/\gamma_{+}^{2}$ is zero. Here, $\gamma_{-}^{2}/\gamma_{+}^{2}=(2-\sqrt{3})/(2+\sqrt{3})\simeq 0.07$. As shown in Eq.~(\ref{generalS}), the snapshot entropy increases as $\gamma_{-}^{2}/\gamma_{+}^{2}$ increases from zero, and reaches a maximum when $\gamma_{+}^{2}=\gamma_{-}^{2}$. The significance of $\gamma_{\pm}$ will be further discussed in the paragraph of the finite-$\chi$ scaling. The third thing is about creation of new fractal images from this algorithm. It is possible to take a reverse process of the present approach to make various beautiful fractals. If $\gamma_{\pm}=1\pm\delta$ in general, the eigenvalues of the single-particle correlator matrix are given by
\begin{eqnarray}
\nu_{\pm}=\frac{1}{2}\pm\frac{\delta}{1+\delta^{2}}.
\end{eqnarray}
Then, changing the $\delta$ value creates a family of various fractal images. We show two examples in Fig.~\ref{fig02}.

Our result suggests that scale-invariance itself is sufficient for the entropy to exhibit a Calabrese-Cardy-type logarithmic scaling. This is reasonable when we remember that the entanglement entropy is logarithm of two point correlation function. The two-point function only with scale invariance is represented as $\left<O_{1}(x_{1})O_{2}(x_{2})\right>=c_{12}/\left(x_{1}-x_{2}\right)^{\Delta_{1}+\Delta_{2}}$ with scaling demensions $\Delta_{1}$ and $\Delta_{2}$, but that of full conformal symmetry has the form $\left<O_{1}(x_{1})O_{2}(x_{2})\right>=c_{12}\delta_{\Delta_{1}\Delta_{2}}/\left(x_{1}-x_{2}\right)^{2\Delta_{1}}$. Although the constraint coming from only scale invariance is somehow weaker than that from the conformal invariance, the form of two-point function does not change so much.

We gain a more intuitive understanding from a holographic perspective reminiscent of the Ryu-Takayanagi formula. The entropy is proportional to the fractal level $N$, as we have already seen in Eq.~(\ref{generalS}). This supports the fact that we have fractal data spanning $N$ different length scales, which can be hierarchically organized into an emergent space with an extra dimension associated with the scale.

One paticular finding from this analysis is that there is no constant correction in Eq.~(\ref{log}). Going back to the previous work on the snapshot of the Ising spin model, we have observed the negative constant contribution to the entropy $S\simeq\ln L -2$ at $T_{c}$, and the origin of this term was unresolved~\cite{Matsueda}. When we take $L\rightarrow 0$, we naively think that there is no information. In the fractal case, this seems to be correct. The Ising spin result reminds us the presence of the negative entropy contribution coming from possible topological effects in the quantum side. Since we are now looking at the holography side, the negative term would come from symmetry breaking, not symmetry protection. Thus, we think that the negative term in the Ising case comes from breaking of $Z_{2}$ zymmetry due to the presence of the ferromagnetic large back ground characterized by $M^{(1)}(x,y)$.

Next, we discuss about finite-$\chi$ scaling. As we have already proven analytically, the singular value spectrum is degenerate with the degeneracy of the $j$-th independent singular value $\bar{\lambda}_{j}$ being the binomial coefficient $\alpha_{j}$. If we focus on the first $(N+1)$-singular values, the coarse-grained snapshot entropy for $1\le\chi\le N+1$ can be represented by
\begin{eqnarray}
S_{\chi}=-\lambda_{1}\ln\lambda_{1}-\left(\chi-1\right)\lambda_{2}\ln\lambda_{2},
\end{eqnarray} 
since $\lambda_{2}=\lambda_{3}=\cdots =\lambda_{N+1}$. Since the entropy becomes zero for the extrapolation $\chi\rightarrow 0$, we obtain
\begin{eqnarray}
-\lambda_{1}\ln\lambda_{1}=-\lambda_{2}\ln\lambda_{2}.
\end{eqnarray}
This leads to the following linear-scaling formula
\begin{eqnarray}
S_{\chi}=S_{1}\chi.
\end{eqnarray}
Therefore, this is quite in contrast to the standard finite-$\chi$ scaling in the quantum 1D near-critical system
\begin{eqnarray}
S_{\chi}=\frac{c\kappa}{6}\ln\chi=\frac{1}{\sqrt{12/c}+1}\ln\chi,
\end{eqnarray}
with the finite-$\chi$ scaling exponent $\kappa$.

\begin{figure}[htbp]
\begin{center}
\includegraphics[width=8cm]{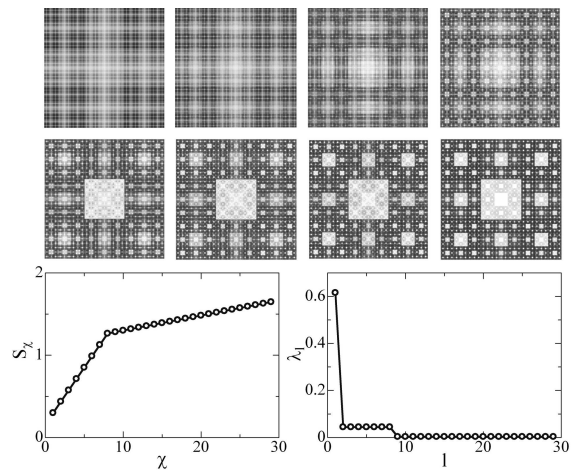}
\end{center}
\caption{(upper panels) $M_{\chi}$ for $\chi=1,2,...,8$, (lower left) Coarse-grained snapshot entropy as a function of $\chi$, (lower right) Singular value spectrum. We find that $\lambda_{1}=0.6155$, $\lambda_{2}=\cdots=\lambda{8}=0.0442$, and $\lambda_{9}=\cdots=\lambda_{29}=0.0032$.}
\label{fig03}
\end{figure}

Figure~\ref{fig03} shows numerical results for $M_{\chi}(x,y)$, $S_{\chi}$, and $\lambda_{l}$ of the white-centered Sierpinski carpet with $L=3^{7}$. It is noted that the hierarchy of the fractal process terminates within a finite level. As shown in Fig,~\ref{fig03}, the entropy is really a linear function with $\chi$, and has kink structure at $\chi=8$. By compaing $M_{\chi}$ with $S_{\chi}$, it is clear that the spatial resolution increases with $\chi$, and at the kink of the entropy we reach at the minimal length scale. Increasing the $\chi$ value thus corresponds to the repetition of the scale transformation. Above the kink, we also observe the linear scaling, and this is also due to the next plateau in the degenerate singular value spectrum. When we observe $M_{\chi}$, the insides of the white regions have somehow bright fine structures. Deeper layers of SVD with large indices, that is information above the kink, remove these extra structures. We also show the singular value spectrum in Fig.~\ref{fig03}. We find that the spectrum is completely degenerate in the linear-$\chi$ scaling region. The number of the degenerate singular values is exactly $\alpha_{N-1}(=7)$ for the first plateau, and $\alpha_{N-2}(=21)$ for the second plateau. The degeneracy is a direct evidence of scale invariance. Therefore, the scaling formula is not logarithmic, but linear due to the scale invariance.

As we have suggested many times, the fractal images lack the full conformal symmetry. We have scale invariance, but do not have Poincare and special conformal symmetries. The conformal transformation does not change the local angle, but horizontal and vertical lines in the image are not kept. This is bad for SVD, since the degeneracy of the spectrum changes completely. In the spin model case at $T_{c}$, however, the global rotation does not change the overall spin structure in the continuous limit. It is curious whether it is possible to physically rotate the spin snapshot of the Ising model and whether we can actually confirm the invariance of the logarithmic scaling after the rotation. Unfortunately, this is technically hard. This is because the physical rotation changes the square-lattice structure of matrix data, and then re-mapping of the rotated data onto a new matrix induces extra entropy that modifies the essential information. This will be a subject of future work.

Summarizing, we have examined the snapshot entropy and spectrum of fractal images. We have mathematically proved and numerically confirmed that the logarithmic scaling of the snapshot entropy agrees with the Calabrese-Cardy formula in 1D free fermions. However, the finite-$\chi$ scaling for the fractal images is different from that in quantum near-critical 1D systems, and this comes from difference between scale and full conformal symmetries. Our conclusion is that the snapshot is a very powerful tool for intuitive understanding of holography, and more detailed examinations will give us meaningful information.

CH is supported by the Agency of Science, Technology and Research of Singapore. HM wrote this paper during his stay in Berkeley and Stanford. HM is greatful for hospitality and comments from Joel Moore and Xiaoliang Qi. HM acknowledges financial support from institute of national colleges of technology, Japan.

\end{document}